\documentclass[a4paper]{JAC2003}
\usepackage{graphicx, epsfig}
\usepackage{booktabs}
\usepackage{varwidth}
\usepackage{xcolor}

\begin{document}
\title{BEAM--BEAM EFFECTS IN SPACE CHARGE DOMINATED ION BEAMS
\thanks{Work supported under Contract Number DE-AC02-98CH10886 with the
auspices of the US Department of Energy}}

\author{C. Montag, A. Fedotov, Brookhaven National Laboratory, Upton, NY, USA}

\maketitle

\begin{abstract}
During low-energy operations below the regular injection energy in the
Relativistic Heavy Ion Collider (RHIC), significant beam lifetime reductions
due to the beam--beam interaction in conjunction with large space charge
tune shifts have been observed. We report on dedicated experiments aimed
at understanding this phenomenon as well as preliminary simulation results,
and propose alternative working points to improve the beam lifetime in
future low-energy RHIC runs.
\end{abstract}

\section{INTRODUCTION}
\begin{figure}
\begin{center}
\epsfig{file=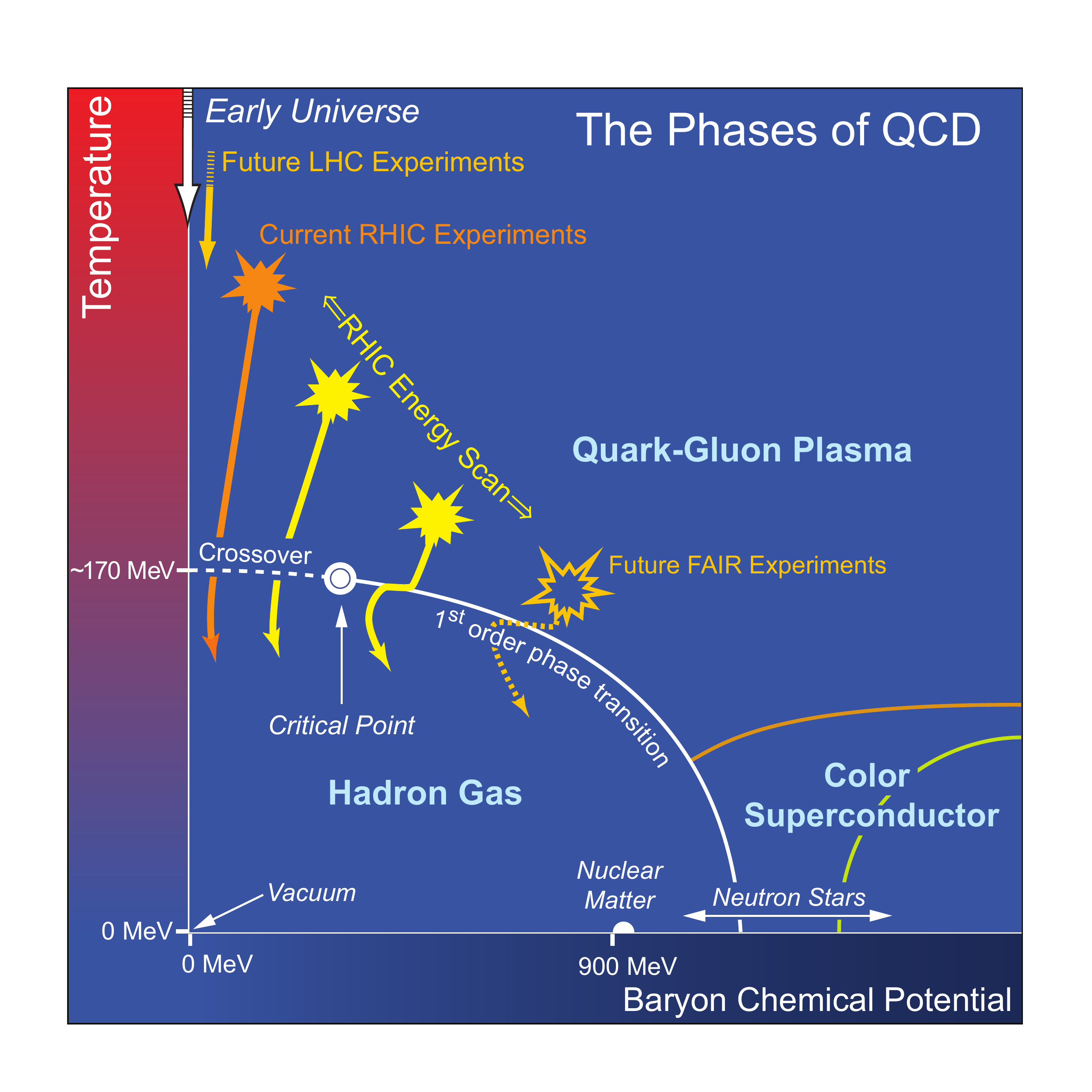, width=\columnwidth}
\end{center}
\caption{\label{qcd}The QCD phase diagram. A lower centre-of-mass energy
$\sqrt{s_{\rm NN}}$ corresponds to a higher baryon chemical potential. The
critical point is expected to be in the energy range between
$\sqrt{s_{\rm NN}}=5$ and $30\,{\rm GeV}$.}
\end{figure}

\begin{figure}
\begin{center}
\epsfig{file=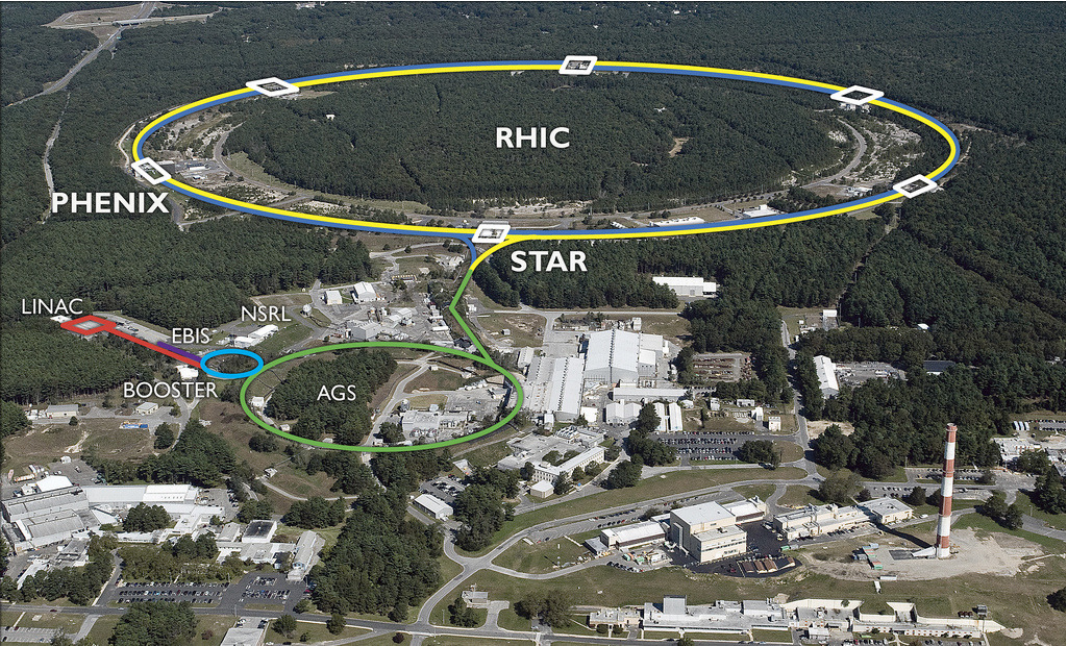, width=\columnwidth}
\end{center}
\caption{\label{rhic}An aerial view of the RHIC accelerator complex, with its two
$3.8\,{\rm km}$ circumference storage rings, `Blue' and `Yellow'.}
\end{figure}

One of the major physics programmes at the RHIC for the next 5--10 years is the
search for the critical point in the Quantum ChromoDynamics (QCD) phase diagram
(Fig.\,\ref{qcd}), which is expected to
occur at centre-of-mass energies in the range of
$\sqrt{s_{\rm NN}}$ = 5--30\,GeV/n.  This requires colliding gold beams with
energies between 2.5 and 15\,GeV/nucleon, which is well below the nominal
energy range of 10--100\,GeV/n in the RHIC (Fig.\,\ref{rhic}).
In conjunction with the circumference
of 3.8\,km, this low energy results in a significant direct space charge tune
shift up to $\Delta Q_{\rm sc}=0.1,$ which is more than ten times larger than the
total beam--beam parameter $\xi_{\rm beam-beam}$ encountered during low-energy
operation \cite{fedotov}.
Experiments with a large beam--beam parameter comparable with the space charge
tune shift have been performed using protons and are reported elsewhere
\cite{fedotov2}.
Table \ref{scandbb} lists space charge and beam--beam
parameters achieved in the RHIC for different Au beam energies.
However, in spite of the large difference in magnitude of these two
effects, and the similar functional dependence of the associated forces on
transverse particle coordinates, we have observed a significant deterioration
of beam lifetimes once beams are brought into collision.

To gain a better understanding of this phenomenon, we have performed a series
of beam experiments and developed a simulation code. In the following
sections, we describe our experimental observations during regular low-energy
operations with Au ions, as well as those dedicated Au beam experiments.
Furthermore, we report
on first results obtained from simulations.
\begin{table}
\caption{\label{scandbb}Beam Lifetimes $\tau$ with and without Collisions
at Different Energies in the RHIC, with the Corresponding Space Charge Tune Shifts
$\Delta Q_{\rm sc}$ and Beam--Beam Parameters $\xi_{\rm beam-beam}$}
\begin{center}
\begin{tabular}{cccc}
\hline\hline
E [GeV/n] & $\Delta Q_{\rm sc}$ & $\xi_{\rm beam-beam}$ & $\tau$ [s]\\
\hline
9.8 & 0.03 & 0 & 2000\\
9.8 & 0.03 & 0.002 & 600\\
\midrule
5.75 & 0.05 & 0 & 1600\\
5.75 & 0.05 & 0.0015 & 400\\
\midrule
5.75 & 0.09 & 0 & 700\\
5.75 & 0.09 & 0.0027 & 260\\
\midrule
3.85 & 0.11 & 0 & 70\\
3.85 & 0.08 & 0.003 & 70\\
\hline\hline
\end{tabular}
\end{center}
\end{table}
\section{EXPERIMENTAL OBSERVATIONS}
\begin{figure}
\begin{center}
\epsfig{file=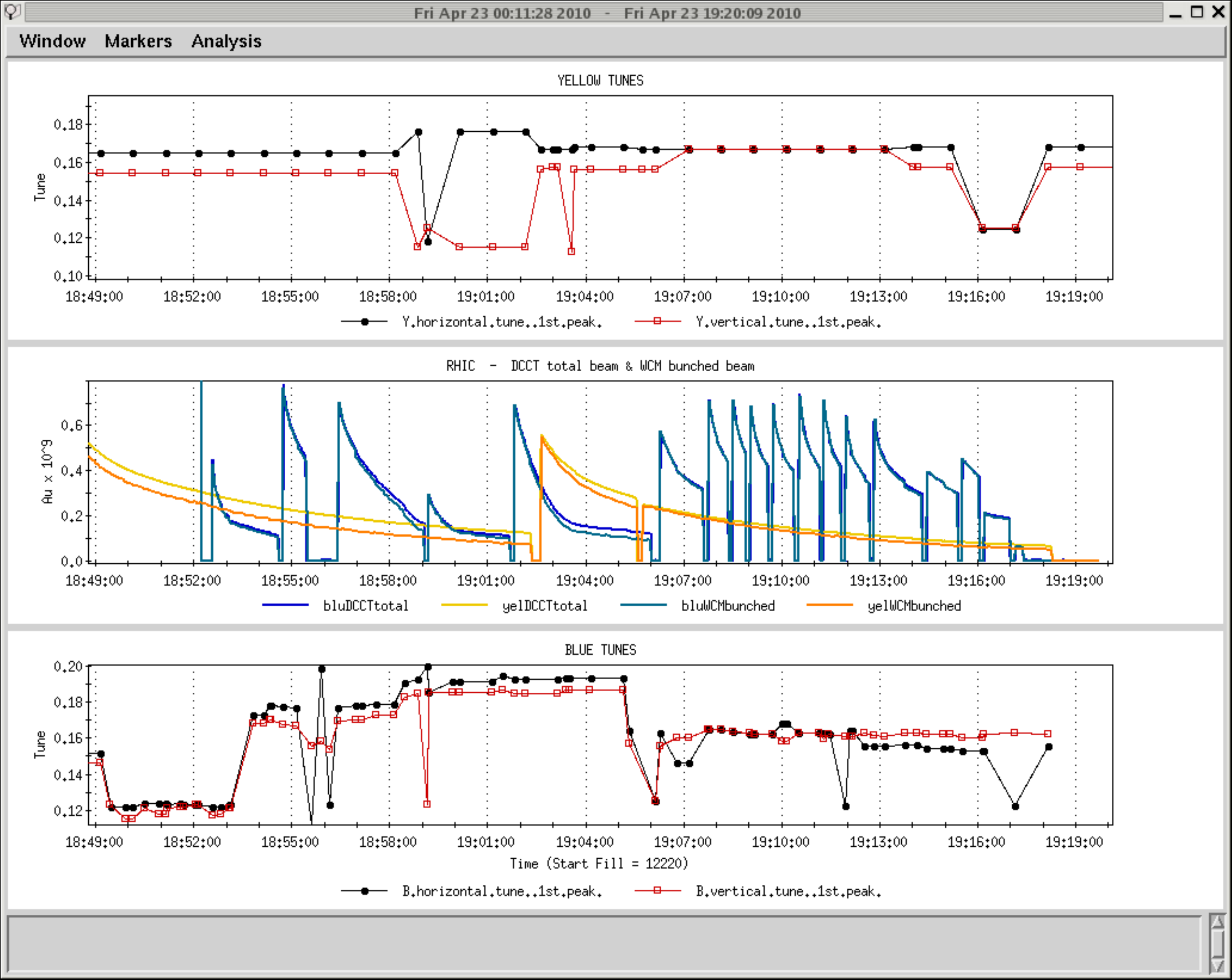, width=\columnwidth}
\end{center}
\caption{\label{lowenergytunescan}Beam intensities (middle plot) in the Blue
and Yellow RHIC rings during a tune scan at $E=5.75\,{\rm GeV/n}$ beam energy.
The Yellow tunes are shown in the top part of the plot and the Blue tunes in the
bottom third. The best Blue beam lifetime is achieved at a working point of
$(Q_x/Q_y)=(28.17/29.16).$}
\end{figure}

When the RHIC was operated at a beam energy of $E=3.85\,{\rm GeV/n},$ a tune
scan was
performed to maximize the beam lifetime. Starting at the regular RHIC heavy
ion working point of $(Q_x/Q_y)=(28.23/29.22),$ the tunes were lowered and the
beam lifetime in collision was observed (see Fig.\,\ref{lowenergytunescan}).
This resulted in a new working point
of $(Q_x/Q_y)=(28.17/29.16)$; during the course of the run this was further
lowered to $(28.13/29.12).$ This latest working point was subsequently used
at $E=5.75\,{\rm GeV/n}$ as well. 

\begin{figure}
\begin{center}
\epsfig{file=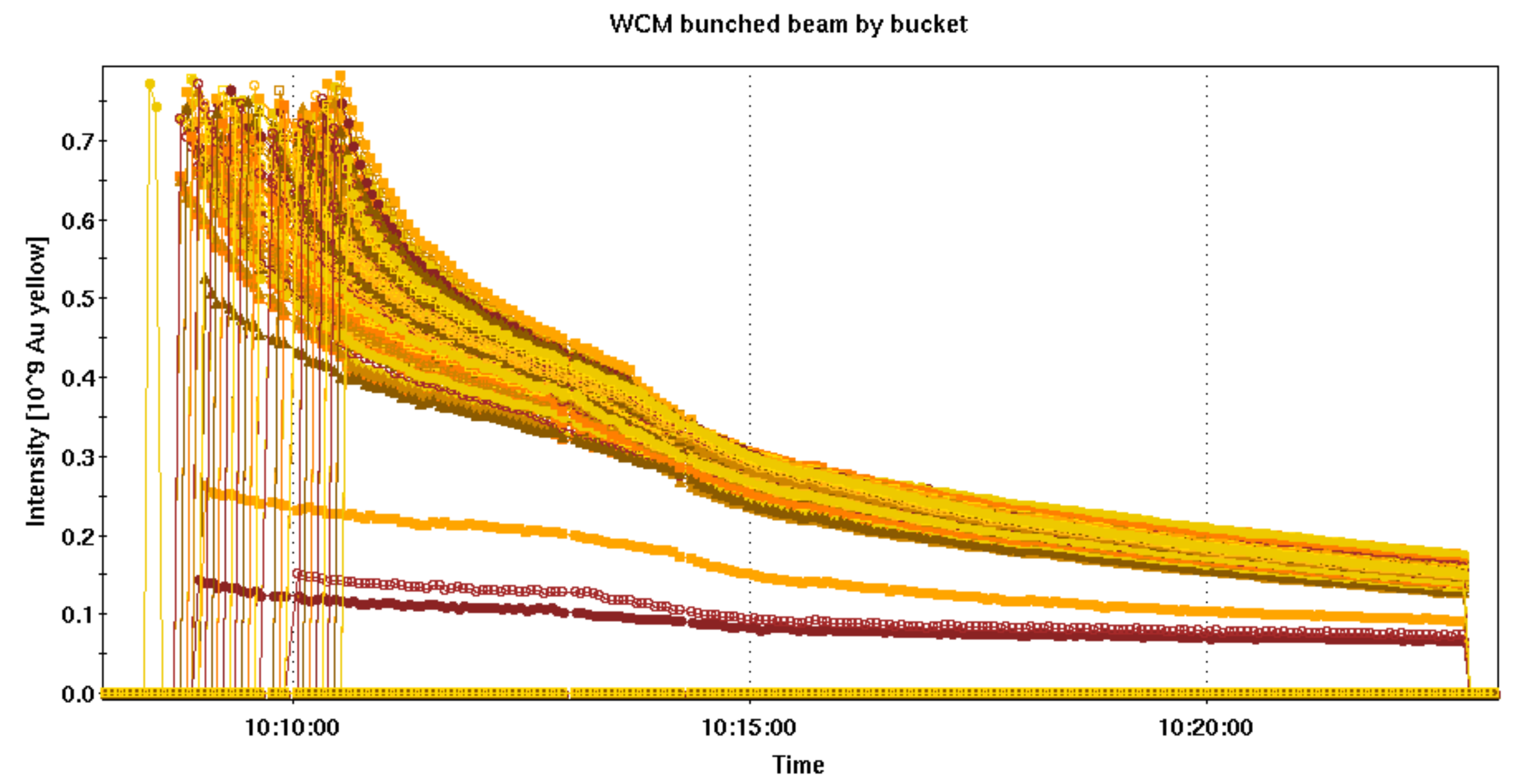, width=\columnwidth}
\end{center}
\caption{\label{store14201wcm}The intensities of individual bunches in the Yellow
ring, at $E=3.85\,{\rm GeV/n}$ beam energy. Collisions start at 10:14, resulting
in a sudden decrease in the lifetime.}
\end{figure}

\begin{figure}
\begin{center}
\epsfig{file=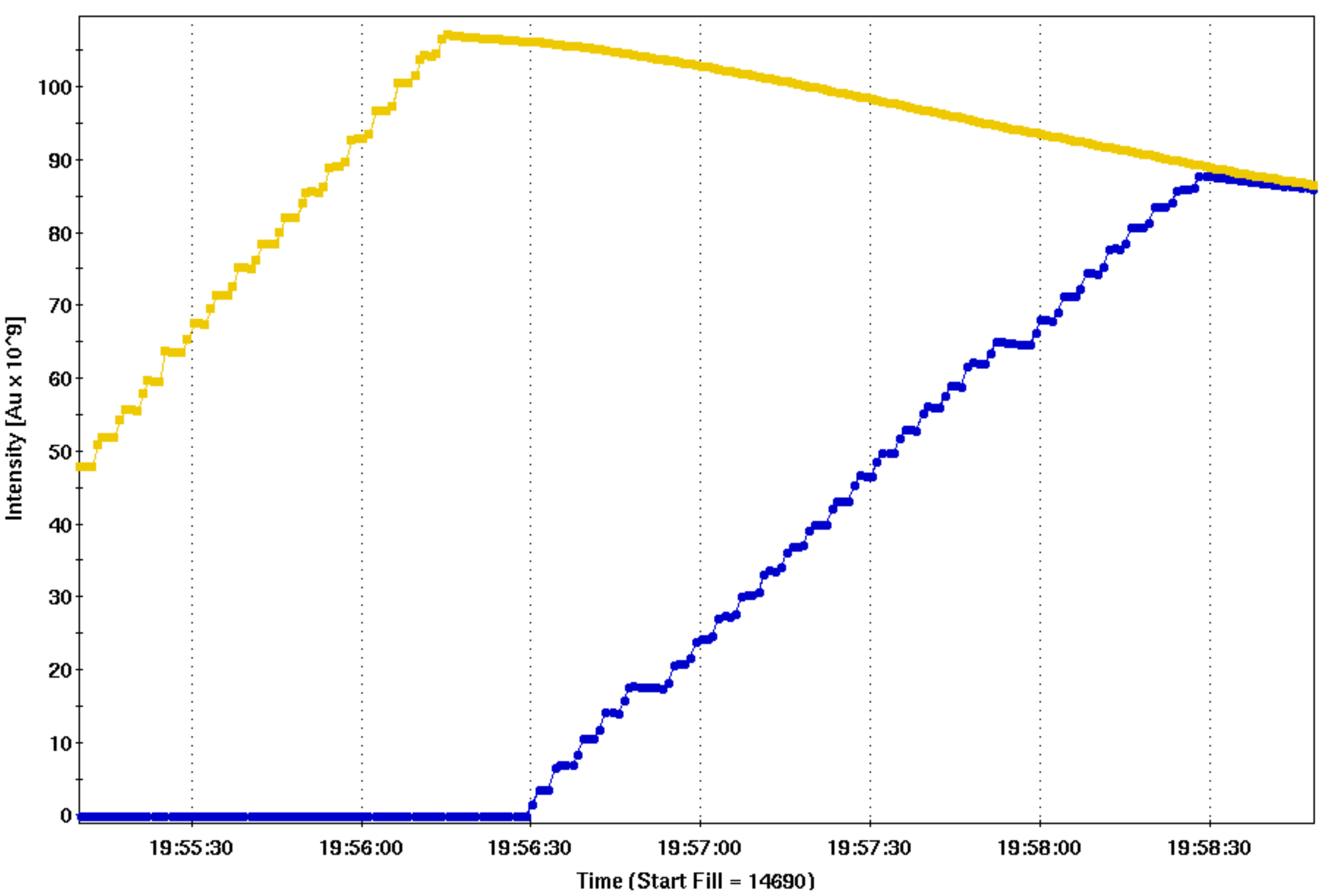, width=\columnwidth}
\end{center}
\caption{\label{beambeamstartofstore}The total beam intensity in the two RHIC
rings during injection, without any transverse separation of the two beams.
After the Yellow ring has been filled, Blue is being injected, resulting in a
gradual decrease of the Yellow beam lifetime due to the beam--beam interaction.}
\end{figure}

\begin{figure}
\begin{center}
\epsfig{file=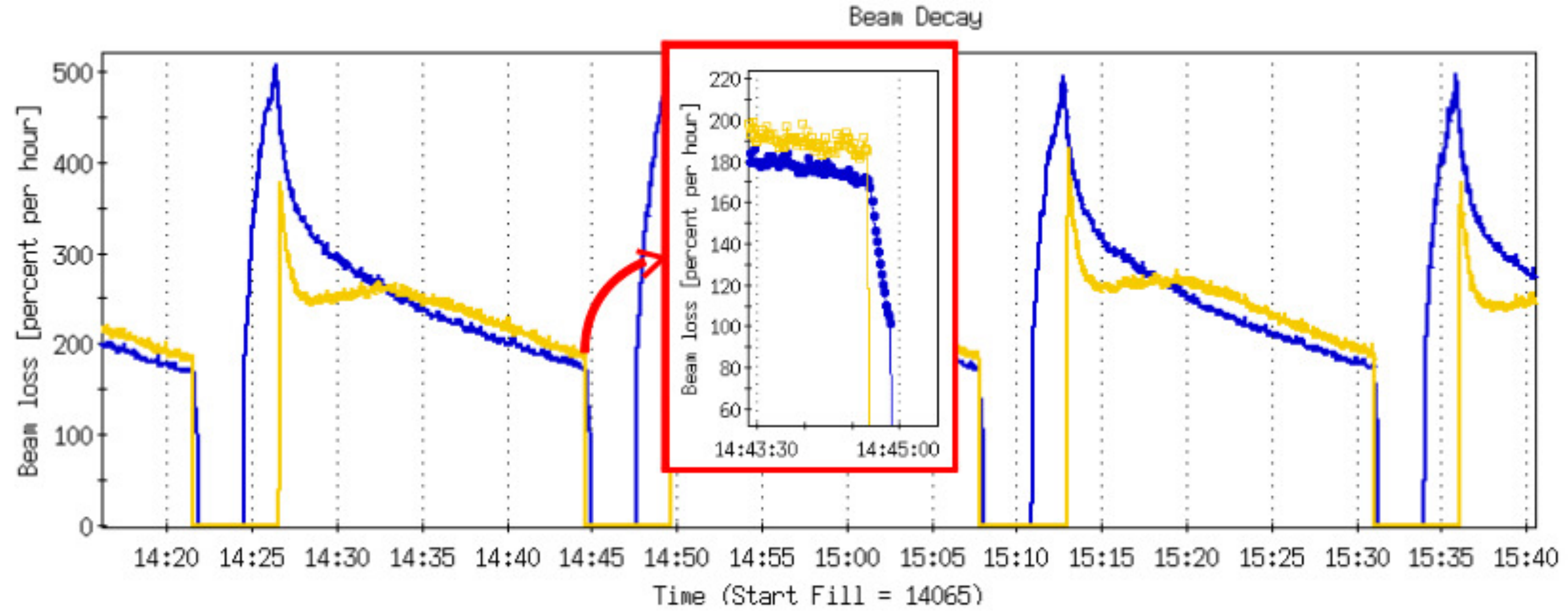, width=\columnwidth}
\end{center}
\caption{\label{beambeam115gev}Beam decay rates during several Au beam stores
at $5.75\,{\rm GeV/n}$ beam energy.
The Blue beam decay rate improves dramatically
as soon as the Yellow beam is dumped at the end of each store (see insert).
Note that the algorithm to calculate the beam decay rate from the measured beam
intensity has a time constant of 20 s.  Hence, the actual drop in the
instantaneous beam decay is even more dramatic than suggested in this picture.}
\end{figure}

During the course of the run, a strong effect of beam--beam interactions on the
lifetime of the space charge dominated beams was consistently observed, as
illustrated in Figs.\,\ref{store14201wcm}--\ref{beambeam115gev}.
Figure\,\ref{store14201wcm} shows the intensity of individual bunches in the
Yellow RHIC ring at a beam energy of $E=3.85\,{\rm GeV/n}.$
Although the initial intensity
drops rather quickly for the first couple of minutes while beams are not
colliding, there is a sudden, sharp decrease in beam lifetime, to roughly the
same level as at the beginning of store, as soon as
the two beams begin colliding. Since the intensity of the Yellow bunch at this time
is only about half the initial value, which reduces the space charge tune shift
by the same factor of 2, this lifetime deterioration cannot simply be explained
by the total tune shift; that is, the sum of the space charge and beam--beam tune shift.
Moreover, a significant beam--beam effect is observed for bunches with a much
smaller intensity, as can be seen in Fig.\,\ref{store14201wcm}.

As a second example, we discuss the evolution of the total intensity of the
Yellow beam at $E=5.75\,{\rm GeV/n}$ while the Blue ring is being filled
(Fig.\,\ref{beambeamstartofstore}). In this case, there is no transverse
separation of the two beams during the injection process, so the injection of
each individual Blue bunch results in a Yellow bunch starting to experience
beam--beam collisions. As a result, the total Yellow beam lifetime slowly
deteriorates the more bunches undergo collisions with the newly injected Blue
bunches.

Finally, we focus on the beam decay rate at the end of a Au store with
$E=5.75\,{\rm GeV/n}$ beam energy (Fig.\,\ref{beambeam115gev}).
When the beam--beam
force on the Blue beam disappears due to dumping of the oncoming Yellow beam,
its decay rate improves dramatically.

Based on operational experience and the desire to improve beam lifetimes
and therefore integrated luminosities in future RHIC low-energy runs, a
dedicated beam experiment aimed at searching for a better working point was
performed. Since the spacing of non-linear resonances is largest in the vicinity
of the integer resonance, fractional tunes below $0.1$ were proposed as the
most promising candidates. These studies were performed at the regular RHIC
injection energy of $9.8\,{\rm GeV/n},$ with a space charge tune shift of
$\Delta Q_{\rm sc}=0.03$ and a beam--beam tune-shift parameter of
$\xi_{\rm beam-beam}=0.002.$

\begin{figure}
\begin{center}
\epsfig{file=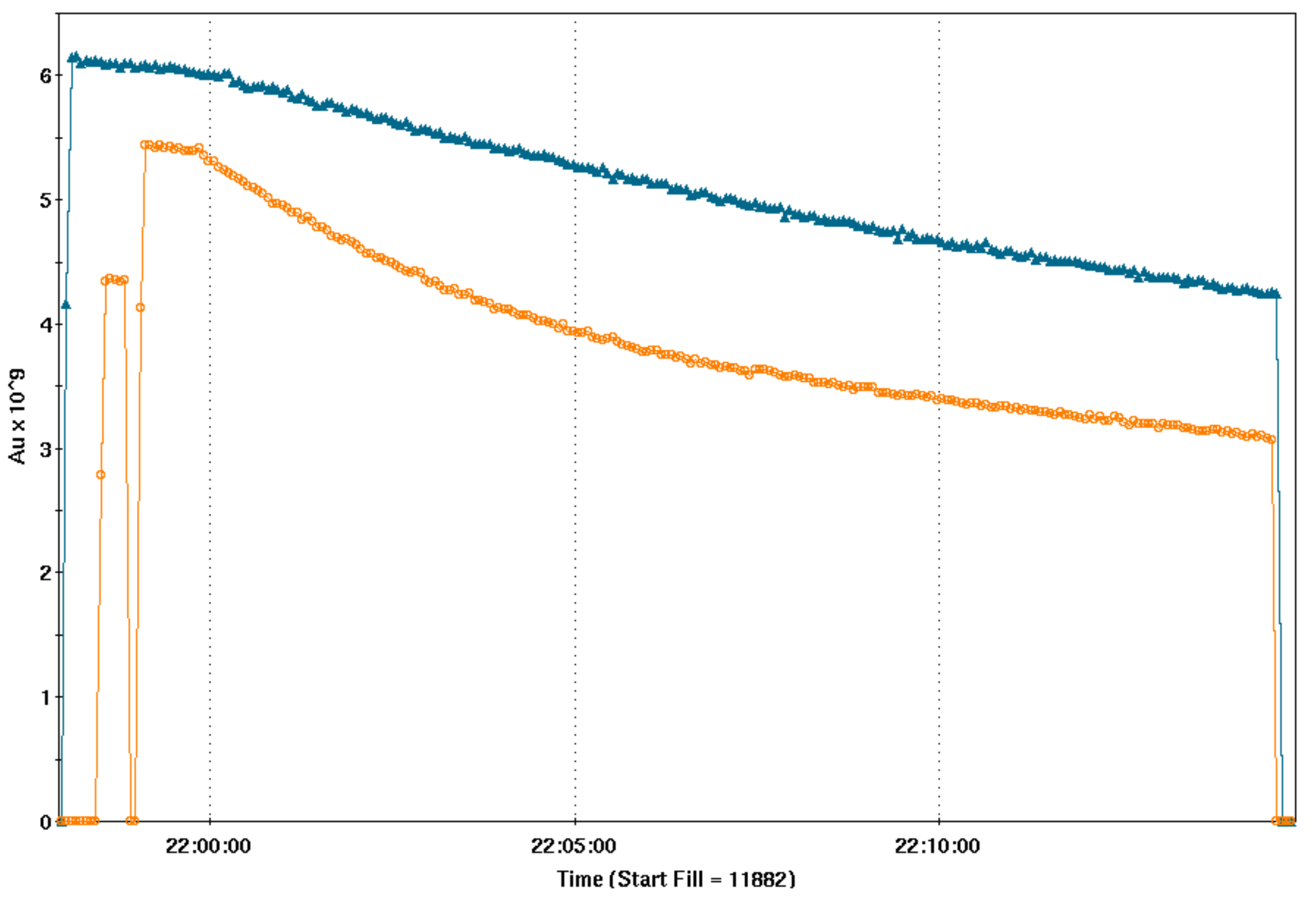, width=\columnwidth}
\end{center}
\caption{\label{injectionenergystore}The beam intensities in the two RHIC rings
during a beam experiment at regular injection energy ($E=9.8\,{\rm GeV/n}$),
at the regular RHIC heavy ion working point of $(Q_x/Q_y)=(28.23/29.22).$
The beams are brought into collision shortly after injection, resulting in
significant decrease in the lifetime.}
\end{figure}

\begin{figure}
\begin{center}
\epsfig{file=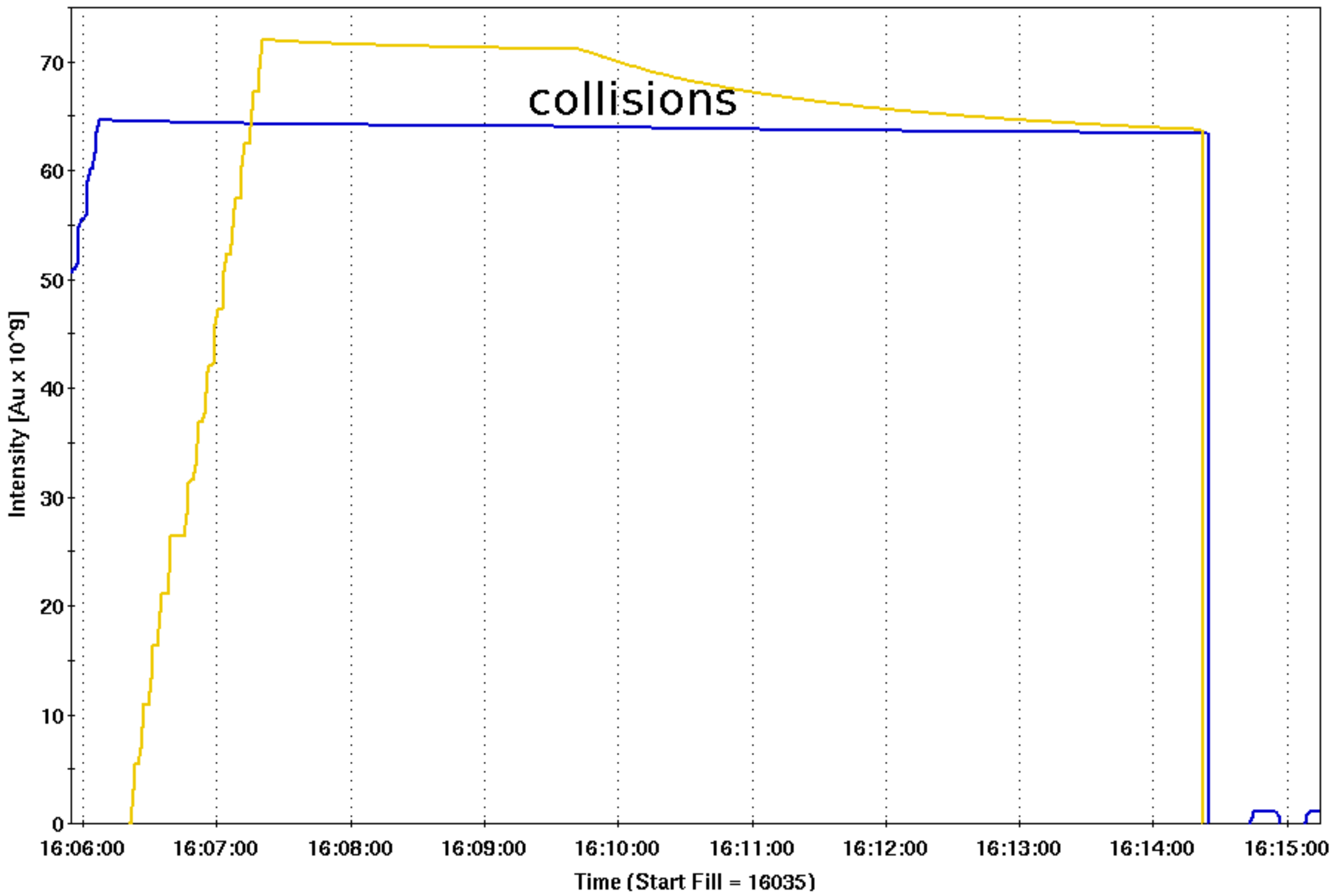, width=\columnwidth}
\end{center}
\caption{\label{beambeamnearinteger}The beam intensities in the RHIC at
$E=9.8\,{\rm GeV/n},$ at a near-integer working point of
$(Q_x/Q_y)=(28.08/29.09)$ in Yellow and $(Q_x/Q_y)=(28.08/29.07)$ in Blue. When
beams are brought into collision, the lifetime of the Yellow beam suffers,
while the Blue beam is unaffected.}
\end{figure}

As already observed during low-energy operations, the beam lifetime deteriorated
substantially when the two beams were brought into collision at the regular
RHIC working point of $(Q_x/Q_y)=(28.23/29.22)$
(Fig.\,\ref{injectionenergystore}).
However, when the experiment was repeated at a near-integer working point
of $(Q_x/Q_y)=(28.08/29.09)$ in Yellow and $(Q_x/Q_y)=(28.08/29.07)$ in Blue,
there was no discernable effect on the Blue lifetime, while the Yellow
lifetime still deteriorated somewhat when beams were brought into collision,
as shown in Fig.\,\ref{beambeamnearinteger}.

The cause of the differing behaviour in the two rings is not yet understood.
It may be attributable to parameters such as chromaticity, coupling control, or
the different working point above the diagonal,
which may have been less than optimal in the Yellow ring during the experiment.
However, this result is very encouraging for future low-energy operations,
although the space charge tune shift during this injection energy experiment
was a factor of 2--3 smaller than at the lower energies.
\section{SIMULATIONS}
To investigate the root cause of the lifetime deterioration, we performed
tracking simulations with a space charge tune shift of $\Delta Q_{\rm sc}=0.06$
and a beam--beam parameter in each of the two RHIC interaction points of
$\xi_{\rm beam-beam}=0.003.$ Using these parameters, tune scans as well
as a frequency map analysis at a fixed working point were applied.
\subsection{The Model}
Space charge simulations are usually very CPU-time consuming because of
frequent recalculations of the particle distribution and the associated
electromagnetic fields. In the particular problem studied here, however, we can
take advantage of the fact that the evolution of the particle distribution
is comparatively slow. This is indicated by the beam lifetime of several
minutes to tens of minutes. Typical simulations track particles only over a
number of turns that corresponds to seconds of real time, so we can safely
assume that the distribution of our test particles does not change appreciably
over the course of the simulation. This approach, which is equivalent to the
weak--strong simulation technique applied in numerical beam--beam studies,
significantly speeds up the computation. In addition, no artificial noise is
introduced into the simulation by the finite number of particles, since
recalculation of the electromagnetic fields from the actual particle distribution
is avoided. Instead, we assume that the distribution remains Gaussian during
the entire simulation process. The r.m.s.\@ width of this Gaussian distribution
is calculated from the beam emittance and the local $\beta$ function, including
the dynamic $\beta$-beat introduced by the space charge and beam--beam forces
around the machine.
The accelerator model is based on the RHIC lattice as described in MAD.  So far,
no lattice non-linearities except the chromaticity correction sextupoles and
the sextupole error in the main dipoles have been included. Particles are
tracked element by element, and space charge kicks are applied at every
quadrupole around the machine. Two beam--beam interaction points are included in
IPs 6 and 8. Synchrotron oscillations are included, and the modulation of the
space charge kick due to the resulting longitudinal position oscillations is
taken into account.
\subsection{Results}
To study the emittance growth as a function of tune, we launch 1000 particles
with a Gaussian distribution in all six phase space coordinates and track them
over 20\,000 turns. At the end of each turn $i$, we calculate the 4-D transverse emittance:
\begin{eqnarray}
\epsilon(i)=\left|
\begin{array}{cccc}
\langle xx\rangle & \langle xx^{\prime}\rangle & \langle xy\rangle & \langle
xy^{\prime}\rangle\\
\langle x^{\prime}x\rangle & \langle x^{\prime}x^{\prime}\rangle & \langle x^{\prime}y\rangle & \langle
x^{\prime}y^{\prime}\rangle\\
\langle yx\rangle & \langle yx^{\prime}\rangle & \langle yy\rangle & \langle
yy^{\prime}\rangle\\
\langle y^{\prime}x\rangle & \langle y^{\prime}x^{\prime}\rangle & \langle
y^{\prime}y\rangle & \langle
y^{\prime}y^{\prime}\rangle\\
\end{array}
\right|^{\frac{1}{4}},
\end{eqnarray}
where $\langle\cdots\rangle$ indicates the average over all particles.
The emittance growth rate
\begin{eqnarray}
\tau_{\epsilon}^{-1}=\frac{1}{\epsilon}\frac{{\rm d}\epsilon}{{\rm d}t}
\end{eqnarray}
is then computed as a function of tune by a linear fit to this 4-D emittance
evolution.

\begin{figure}
\begin{center}
\epsfig{file=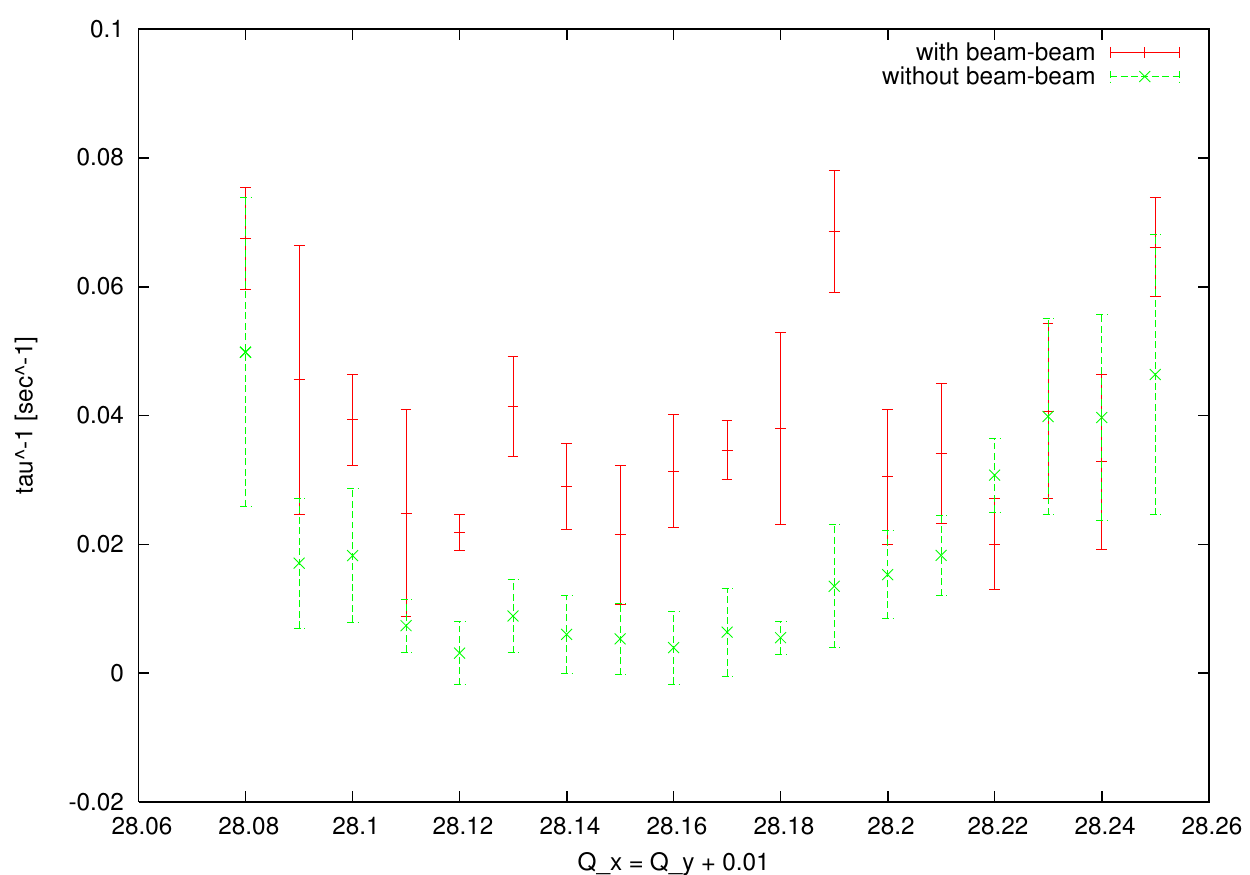, width=\columnwidth}
\end{center}
\caption{\label{tunescan_1312}The 4-D emittance growth rate
$\tau_{\epsilon}^-1=\frac{1}{\epsilon}\frac{{\rm d}\epsilon}{{\rm d}t}$ as
a function of tune, with the working point $(Q_x/Q_y)$ chosen such that
$Q_y=Q_x-0.01.$}
\end{figure}

For the initial tune scan, depicted in Fig.\,\ref{tunescan_1312},
we varied the tunes in steps of
$\Delta Q_{x,y}=0.01,$ with $Q_y=Q_x-0.01.$ In the absence of the beam--beam
interaction, the 4-D emittance growth rate at fractional tunes below 0.2 is
significantly lower than above, which qualitatively agrees with experimental
observations. Once beam--beam interactions are added, the emittance growth rate
increases over most of the tune range; however, the resulting growth rate
is, within error bars, independent of tune.

To determine the tune footprint and the tune diffusion frequency, map analysis
\cite{laskar1, laskar2} was
applied at a fixed working point of $(Q_x/Q_y)=(28.13/29.12)$ for the two cases
with and without beam--beam interaction. For this purpose, we track a single test
particle over $2^{14}$ turns and apply fast Fourier transforms to calculate
the horizontal and vertical tunes $(Q_{x,1}/Q_{y,1})$ and $(Q_{x,2}/Q_{y,2})$
for the first and second $2^{13}$ turns. To increase the tune resolution, we
apply an interpolation technique \cite{bazzani}.

\begin{figure}
\begin{center}
\epsfig{file=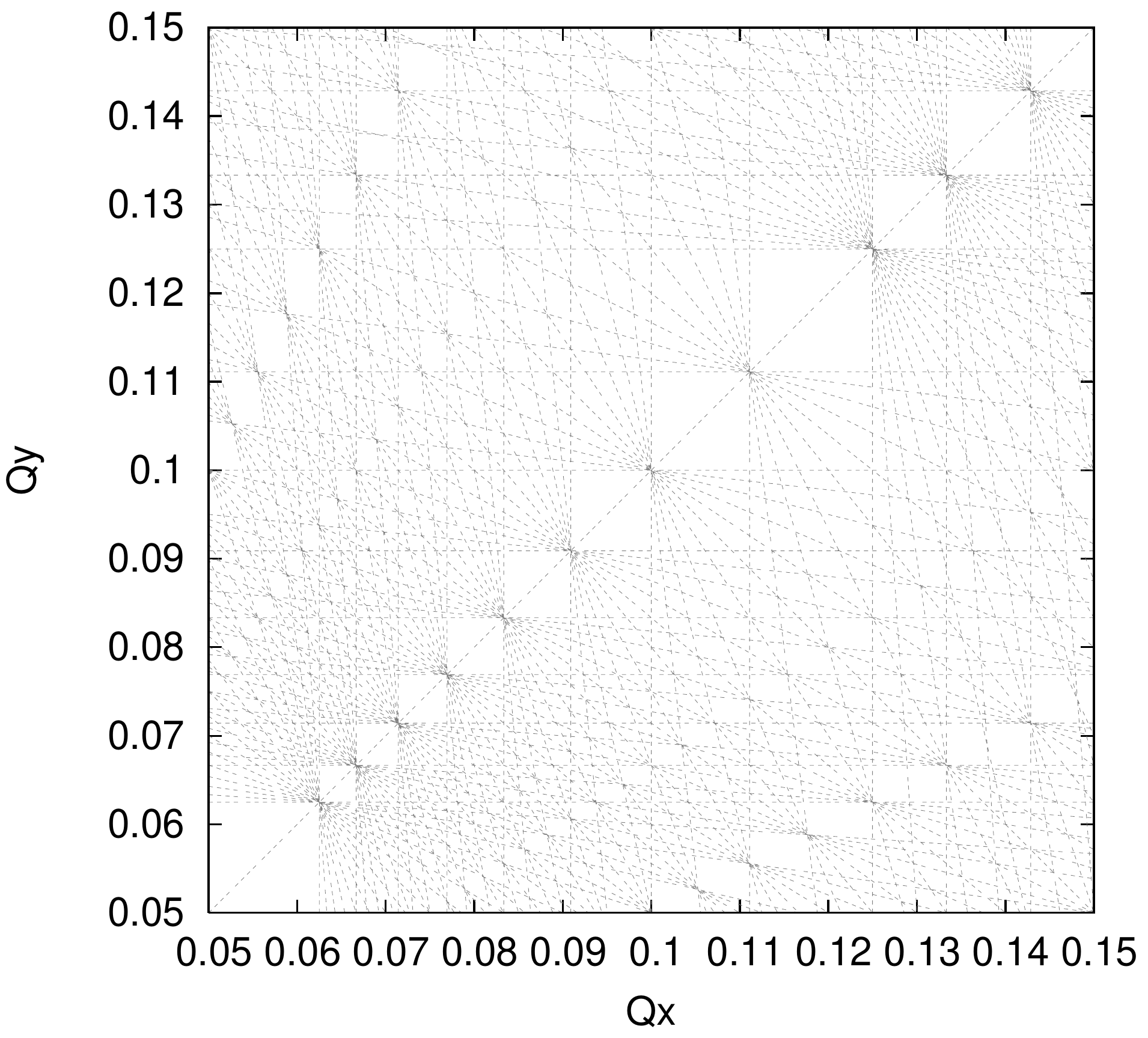, angle=00, width=\columnwidth,
height=0.8\columnwidth}
\epsfig{file=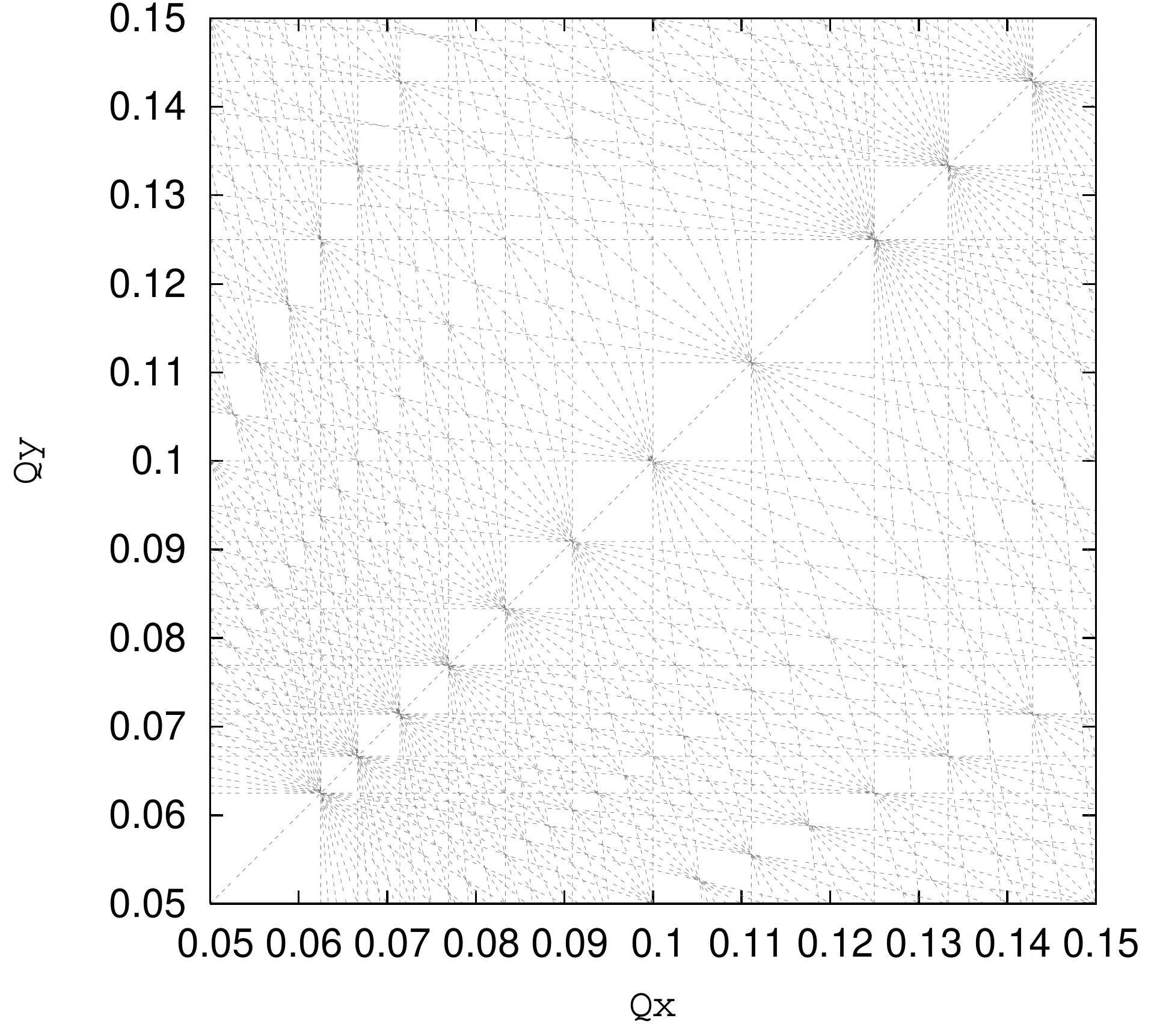, angle=00, width=\columnwidth,
height=0.8\columnwidth}
\end{center}
\caption{\label{fma_1312}Tune footprints at a nominal working point of
$(Q_x/Q_y)=(28.13/29.12),$ without (top) and with (bottom) beam--beam
interaction.}
\end{figure}

The tune diffusion is measured as
\begin{eqnarray}
|\Delta Q|=\sqrt{|Q_{x,1}-Q_{x,2}|^2+|Q_{y,1}-Q_{y,2}|^2}.
\end{eqnarray}
The resulting tune footprint and tune diffusion is plotted in
Fig.\,\ref{fma_1312}. While the tune footprint overlaps the coupling resonance
$Q_x=Q_y$ with as well as without beam--beam interaction, the presence of the
beam--beam force significantly enhances the tune diffusion around that resonance.

\begin{figure}
\begin{center}
\epsfig{file=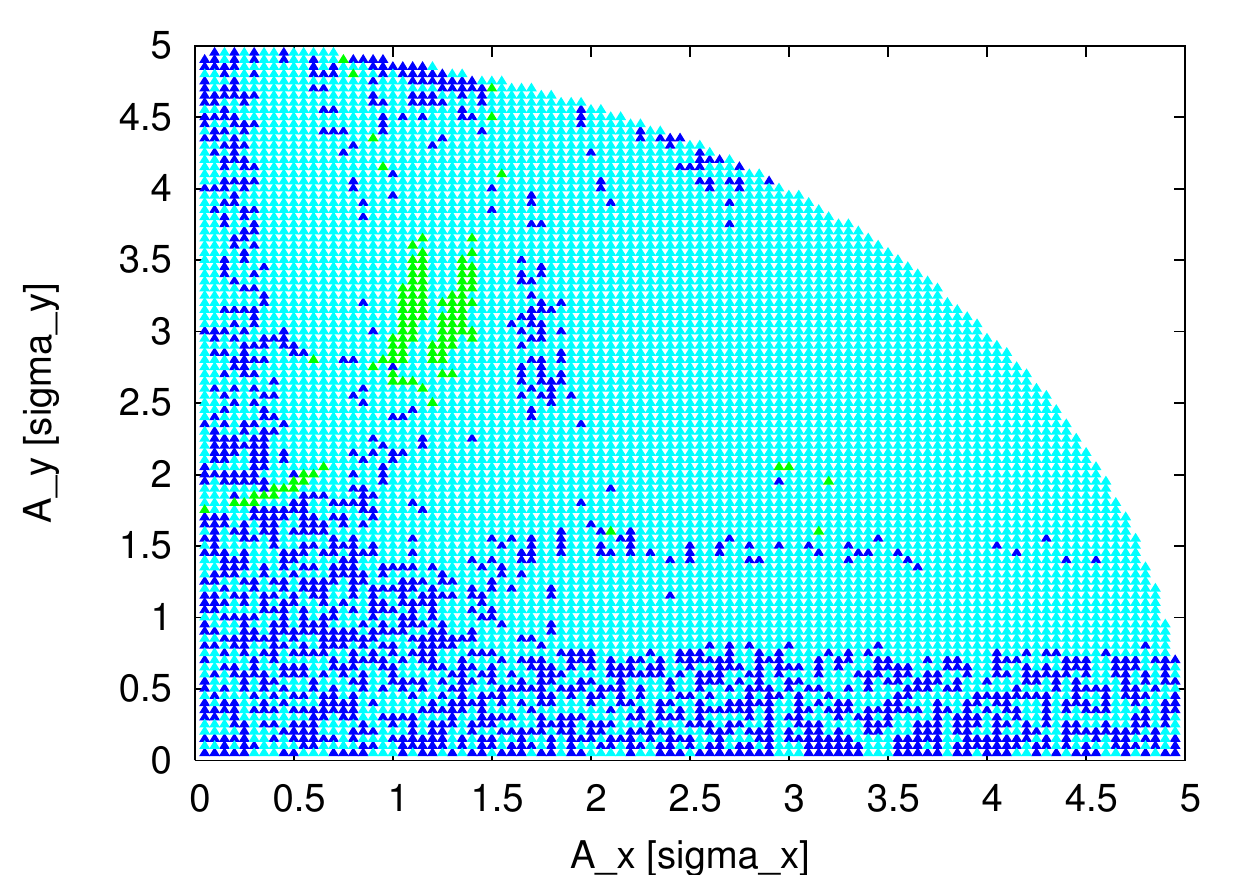, width=\columnwidth,
height=0.8\columnwidth}
\epsfig{file=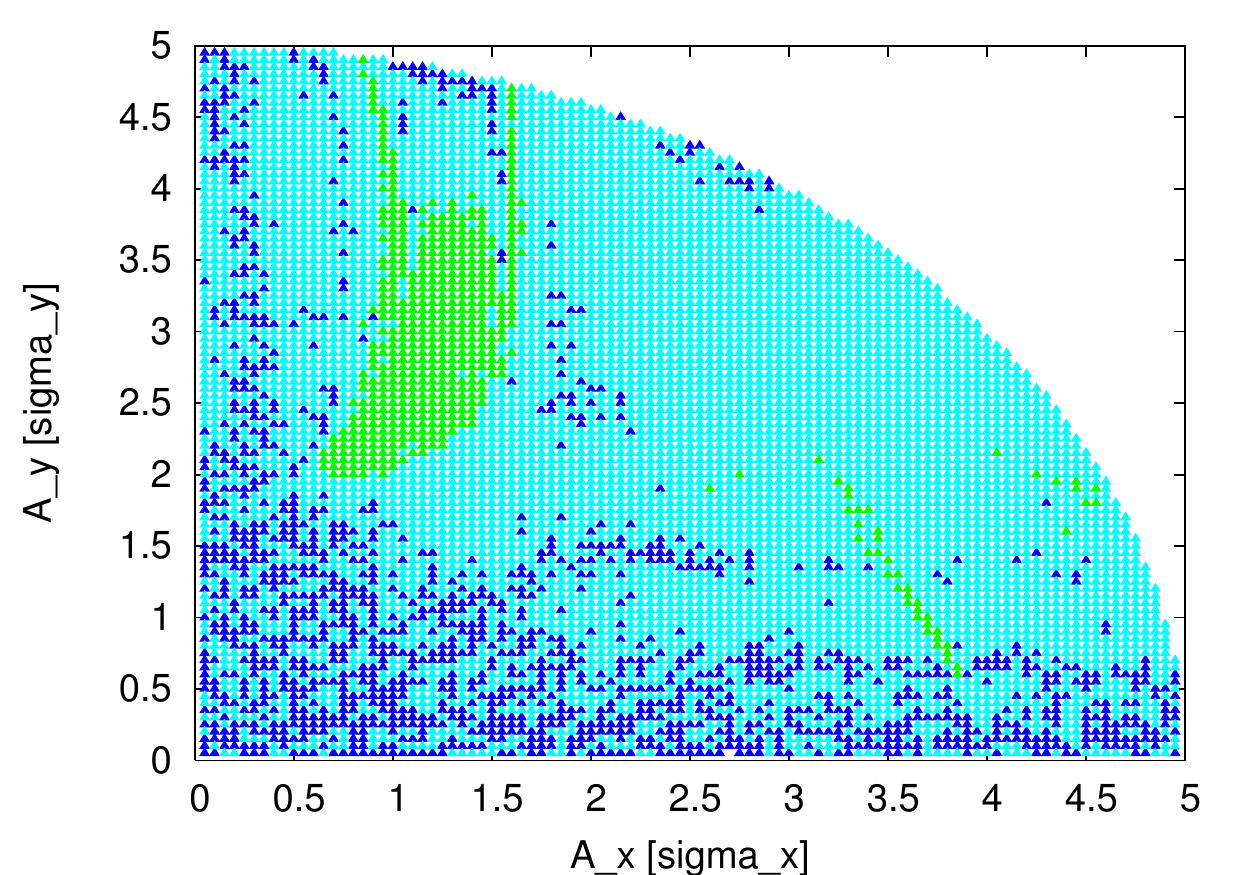, width=\columnwidth,
height=0.8\columnwidth}
\end{center}
\caption{\label{fma_1312_amp}Tune diffusion in the amplitude space, as
obtained from the frequency map analysis at a nominal working point
of $(Q_x/Q_y)=(28.13/29.12),$ without (top) and with (bottom) beam--beam
interaction.}
\end{figure}

Plotting the same data in the amplitude space (Fig.\,\ref{fma_1312_amp})
reveals that this enhanced tune diffusion occurs for amplitudes $(A_x, A_y)$
in the region $\sigma_x<A_x<2\sigma_x$ and $2\sigma_y<A_y<4\sigma_y.$

\begin{figure}
\begin{center}
\epsfig{file=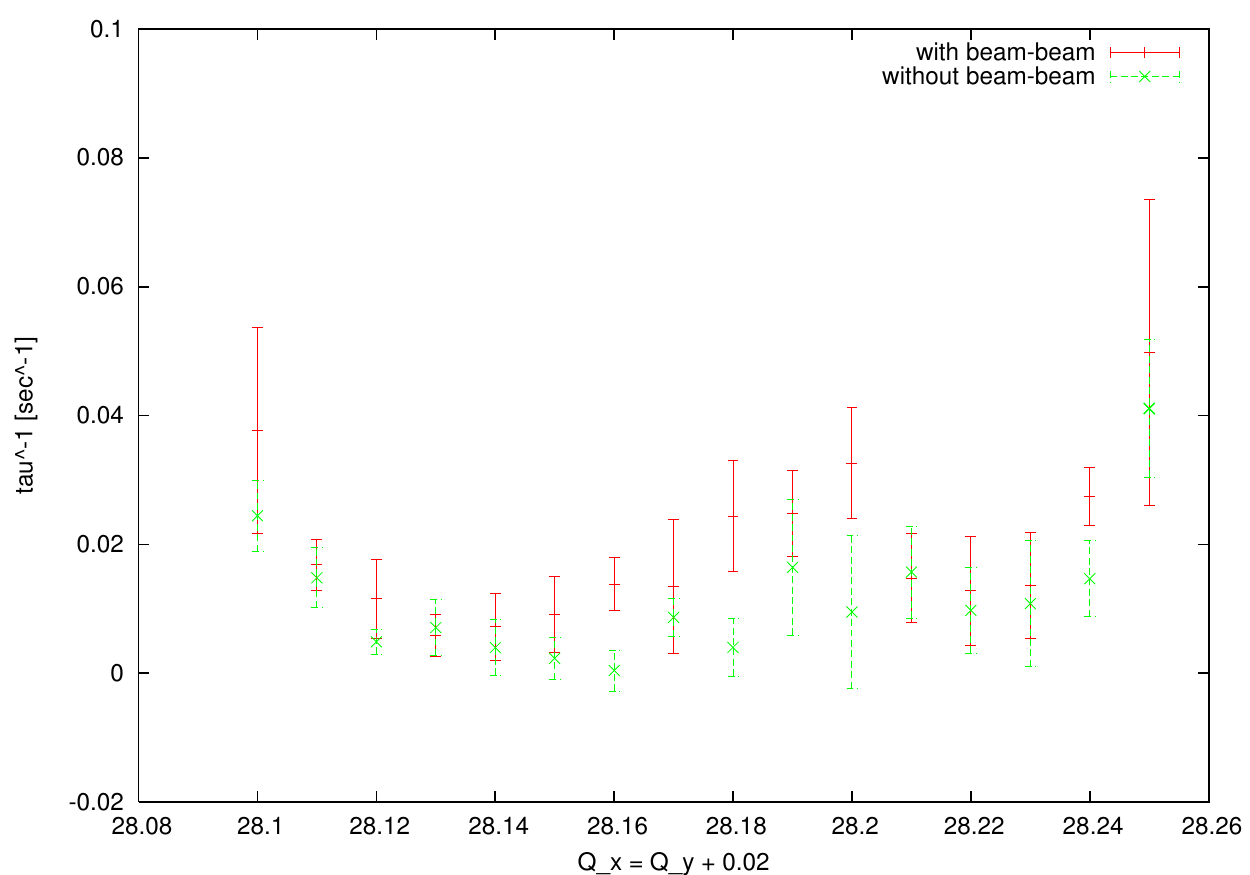, width=\columnwidth}
\end{center}
\caption{\label{tunescan_1412}The 4-D emittance growth rate
$\tau_{\epsilon}^-1=\frac{1}{\epsilon}\frac{{\rm d}\epsilon}{{\rm d}t}$ as
a function of tune, with the working point $(Q_x/Q_y)$ chosen at an increased
tune split with $Q_y=Q_x-0.02.$}
\end{figure}

As shown above, the largest tune diffusion occurs around the coupling
resonance. This behaviour suggests that it might be beneficial to increase the
tune split between the two planes, thus selecting a working point further away
from the coupling resonance. To study this hypothesis, we performed a tune scan
with $Q_y=Q_x-0.02$ and determined the 4-D emittance growth rates. As shown in
Fig.\,\ref{tunescan_1412}, the effect of the beam--beam interaction on
the emittance growth rate is significantly reduced. 
\begin{figure}
\begin{center}
\epsfig{file=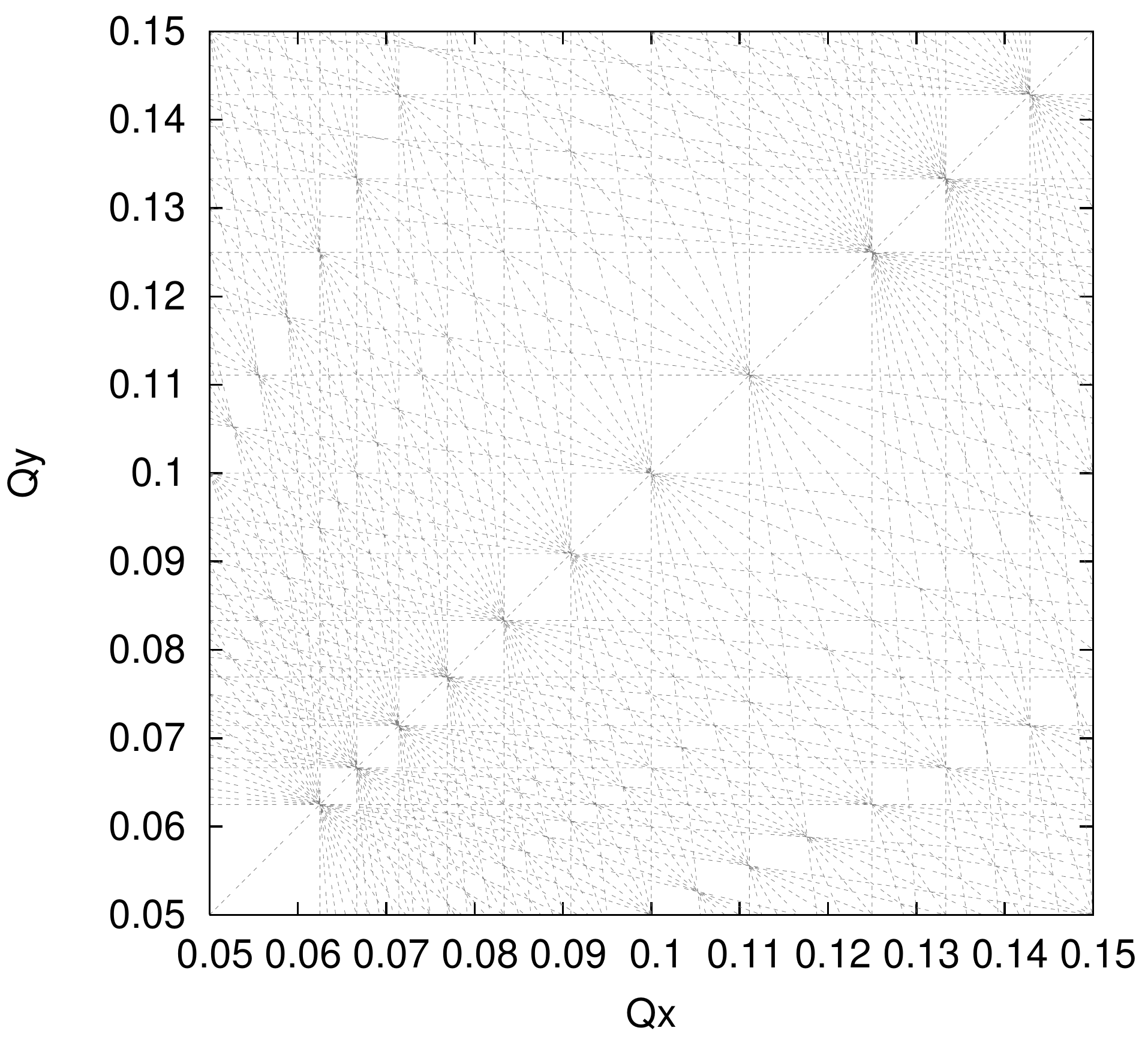, angle=00, width=\columnwidth,
height=0.75\columnwidth}
\epsfig{file=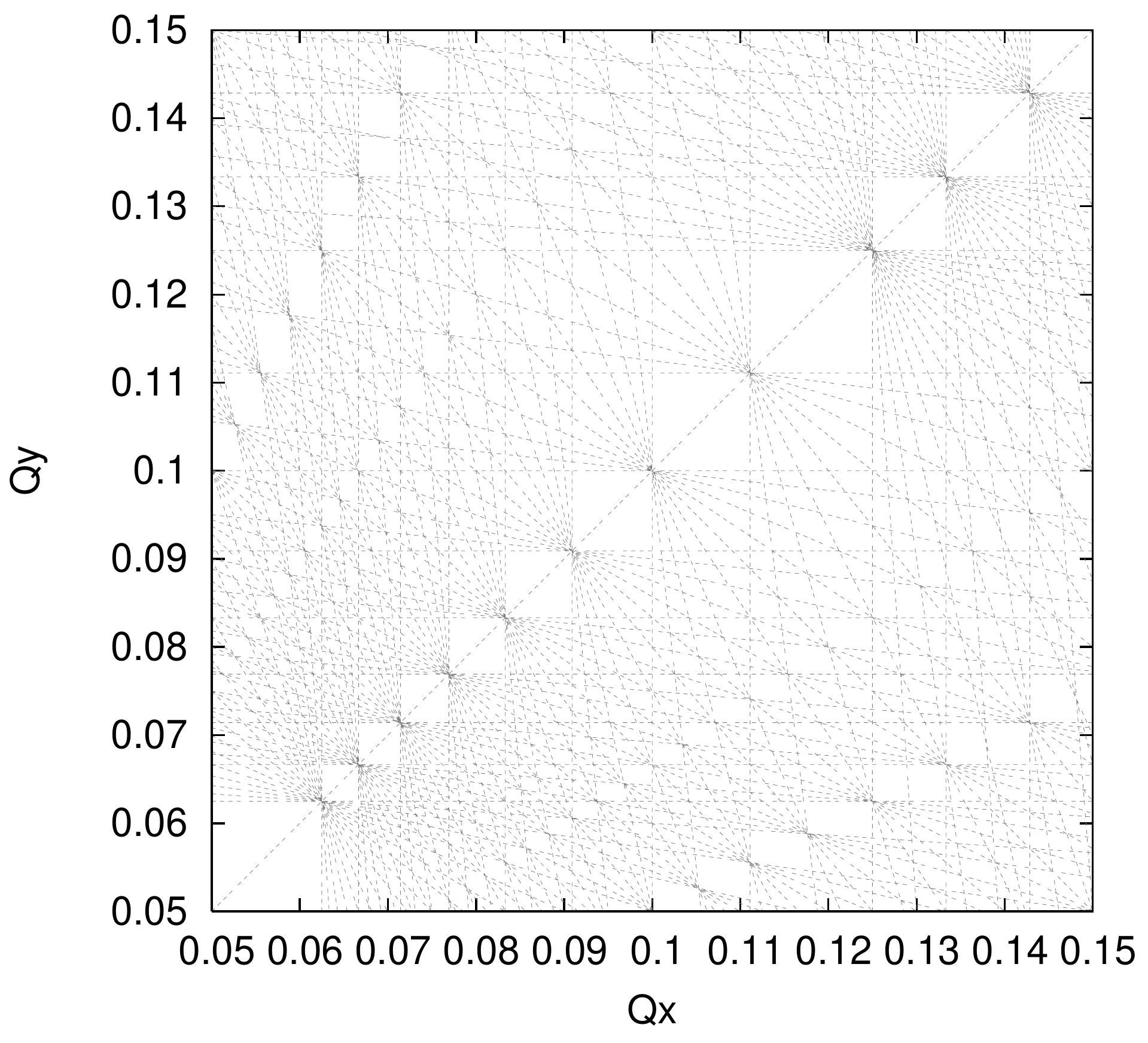, angle=00, width=\columnwidth,
height=0.75\columnwidth}
\end{center}
\caption{\label{fma_1412}Tune footprints at a nominal working point with
an increased tune split between the two planes,
$(Q_x/Q_y)=(28.14/29.12),$ without (top) and with (bottom) beam--beam
interaction.}
\end{figure}
\begin{figure}
\begin{center}
\epsfig{file=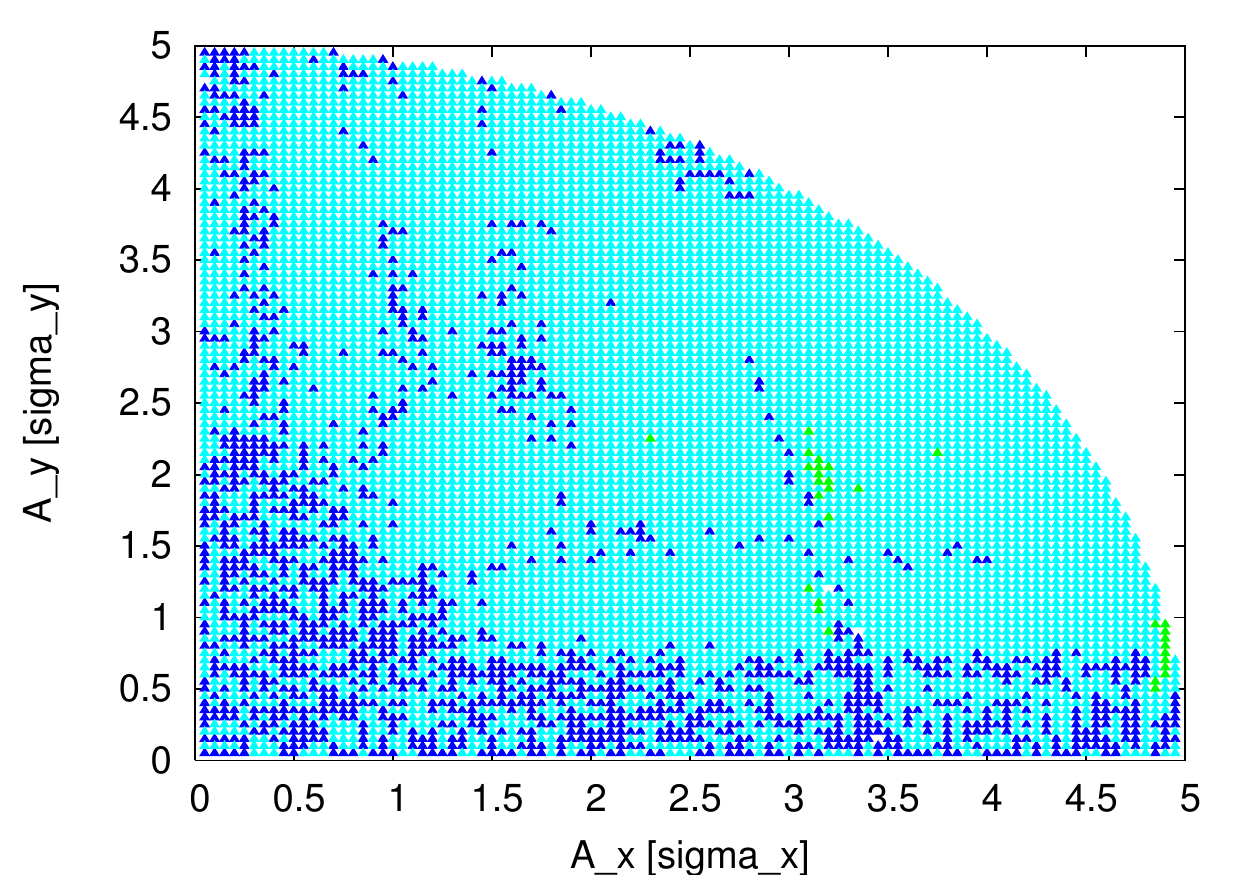, width=\columnwidth,
height=0.75\columnwidth}
\epsfig{file=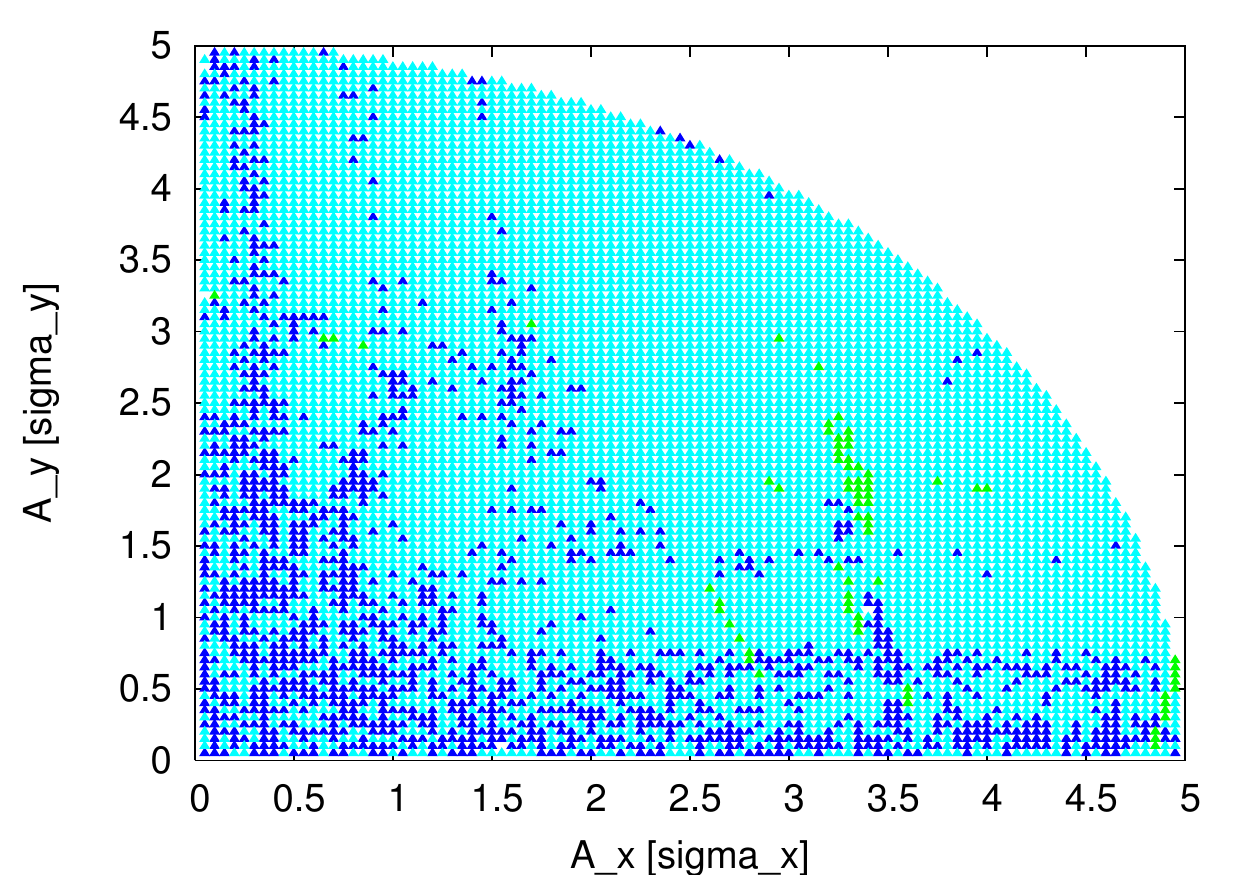, width=\columnwidth,
height=0.75\columnwidth}
\end{center}
\caption{\label{fma_1412_amp}Tune diffusion in the amplitude space, as
obtained from the frequency map analysis at a nominal working point with
an increased tune split between the two planes,
$(Q_x/Q_y)=(28.14/29.12),$ without (top) and with (bottom) beam--beam
interaction.}
\end{figure}

This observation is supported by the results of a frequency map analysis for
the working point $(Q_x/Q_y)=(28.14/29.12)$
(Figs.\,\ref{fma_1412} and \ref{fma_1412_amp}), which shows little effect
of the beam--beam interaction on tune diffusion.

While this result is certainly very encouraging in terms of improving the
machine performance, it is worthwhile repeating that our model so far does
not include any magnet non-linearities beyond sextupoles, which may lead
to increased tune diffusion around the associated non-linear resonances in the
presence of the beam--beam interaction.
\section{Summary}
We have studied the effects of beam--beam interactions in colliding beams with
large direct space charge parameters up to $\Delta Q_{\rm sc}=0.1$ both
experimentally and through simulations. During RHIC low-energy operations as
well as dedicated experiments, we have consistently observed a strong effect of
the beam--beam interaction on the lifetime of the stored beam, although the
associated beam--beam parameter was about an order of magnitude smaller than
the space charge tune shift.

To provide the maximum tune space between non-linear resonances, we have operated
the RHIC at a near-integer working point. In this case, we observed no discernable
lifetime reduction in the Blue ring, while Yellow still suffered. The root
cause of this difference between the two rings is still unknown, and may be
related to parameters such as chromaticity and coupling control, or the
particular working point during the
experiment. However, this result is very encouraging for future low-energy
runs, although the corresponding space charge tune shift of
$\Delta Q_{\rm sc}=0.03$ was comparably modest.

Although a quantitative comparison of our simulation results with experimental
observations in the RHIC is difficult due to the lack of effects such as
intra-beam scattering in the simulation code, the tracking model presented here
qualitatively reproduces the main experimental result, namely the strong effect
of the beam--beam interaction in the presence of a large space charge tune
shift. Based on these simulation results, an alternative working point further
away from the coupling resonance $Q_x=Q_y,$ which appears to be the main
source of emittance growth in our simulations, may be beneficial. However, it
is of utmost importance to repeat these simulations after higher-order multipole
errors have been added to the tracking model, which may lead to a further
reduction of the usable tune space.

\end{document}